\documentclass[journal=apchd5,manuscript=letter,layout=twocolumn]{achemso}

\setkeys{acs}{keywords=true} 
\usepackage{amsmath,amssymb,latexsym}    
\usepackage{wasysym}
\usepackage{amsfonts}
\usepackage{color}
\usepackage{nicefrac}
\usepackage[bookmarks=false,pdffitwindow=true,colorlinks,linkcolor=black,citecolor=black,filecolor=black,urlcolor=blue,pdfborder={0 0 0}]{hyperref}
\usepackage[english]{babel}

\title{Second Harmonic Generation from Critically Coupled Surface Phonon Polaritons}

\author{Nikolai Christian Passler}
\author{Ilya Razdolski}
\author{Sandy Gewinner}
\author{Wieland Sch\"ollkopf}
\author{Martin Wolf}
\author{Alexander Paarmann}
\email{alexander.paarmann@fhi-berlin.mpg.de}
\affiliation{Fritz-Haber-Institut der Max-Planck-Gesellschaft, Faradayweg 4-6,
14195 Berlin, Germany}


\begin{document}

\date{\today}
						

\begin{abstract}
Mid-infrared nanophotonics can be realized using sub-diffractional light localization and field enhancement with surface phonon polaritons in polar dielectric materials. We experimentally demonstrate second harmonic generation due to the optical field enhancement from critically coupled surface phonon polaritons at the 6H-SiC-air interface, employing an infrared free-electron laser for intense, tunable, and narrow-band mid-infrared excitation. Critical coupling to the surface polaritons is achieved using a prism in the Otto geometry with adjustable width of the air gap, providing full control over the excitation conditions along the polariton dispersion. The calculated reflectivity and second harmonic spectra reproduce the full experimental data set with high accuracy, allowing for a quantification of the optical field enhancement. We also reveal the mechanism for low out-coupling efficiency of the second harmonic light in the Otto geometry. Perspectives on surface phonon polariton-based nonlinear sensing and nonlinear waveguide coupling are discussed.
\end{abstract}

\maketitle

\section{Introduction}

Surface polaritons are the key building block of nanophotonics since these excitations allow for extreme light localization accompanied with significant enhancement of the local optical fields. A large body of research has focused on surface plasmon polaritons (SPPs) at noble metal surfaces,\cite{Maier2007} which has led to a number of applications ranging from optical near-field microscopy to nonlinear plasmonic nanosensors.\cite{Sonntag2014, Muller2016,Anker2008,Kabashin2009,Martin-Becerra2013,Mesch2016}  Recently, an alternative approach was introduced employing surface phonon polaritons (SPhPs) which can be excited in the mid-infrared (mid-IR) at the surface of polar dielectrics.\cite{Falge1973,Huber2005,Neuner2009} In these materials, optical phonon resonances in the dielectric response result in a negative permittivity in the reststrahl range between the transverse optical (TO) and longitudinal optical (LO) phonon frequencies and, in consequence, the existence of a highly dispersive surface polariton.\cite{Adachi1999} Notably, the much reduced optical losses of SPhPs as compared to SPPs have been argued to potentially solve the loss problem that was identified as the key limitation for wide-spread implementation of plasmonic devices.\cite{Khurgin2015,Caldwell2014a} Recent pioneering experiments have employed SPhPs for optical switching,\cite{Li2016} as well as sub-diffractional light confinement\cite{Wang2013a,Caldwell2013a,Gubbin2016} and strongly enhanced nonlinear response in sub-wavelength nanostructures.\cite{Razdolski2016}

A systematic study of the linear and nonlinear-optical response of surface polaritons is enabled by prism coupling either in Kretschmann-Raether\cite{Raether1988} or Otto\cite{Otto1968} configuration. In both cases, the high refractive index of the prism operated in a regime of total internal reflection provides the large momenta required to excite the surface waves.\cite{Raether1988} Whilst mostly employed for studies of SPPs,\cite{Palomba2008,Grosse2012,Temnov2012,FoleyIV2015,Razdolski2016b} a few works also investigated SPhPs with prism coupling in the mid-IR.\cite{Falge1973,Okada2008,Neuner2009,Zheng2017} Specifically in the Otto configuration, the surface polariton is excited across an air gap of adjustable width, providing extrinsic tunability of the excitation efficiency which can lead to critical coupling conditions.\cite{Neuner2009} Strong optical field enhancement at critical coupling was predicted but could not be experimentally confirmed using linear optical techniques. Instead, nonlinear-optical approaches like second harmonic generation (SHG) are highly sensitive to the localized electromagnetic fields.\cite{Quail1983,Kauranen2012,Grosse2012,Paarmann2016,Razdolski2016,Razdolski2016b}
      
In this Letter, we experimentally demonstrate the first SHG from critically coupled SPhPs, allowing us to accurately determine the associated optical field enhancement. We employ the Otto geometry for prism coupling to SPhPs at the 6H-SiC-air interface across a variable air gap, and detect reflectivity and SHG output spectroscopically at various positions in the SPhP dispersion, using an infrared free-electron laser (FEL) for tunable narrowband excitation. By varying the air gap width between the prism and the sample, we demonstrate critical coupling behavior of the SPhP excitation efficiency. The calculated linear and nonlinear response reproduces the full dataset of reflectivity and SHG spectra with high accuracy. We extract the optical field enhancement, analyze the out-coupling of the SHG intensity in the Otto geometry, and discuss several potential applications of the nonlinear response from SPhPs.


\section{Results and Discussion}

\begin{figure*}[htb]
\includegraphics[width = .9\textwidth]{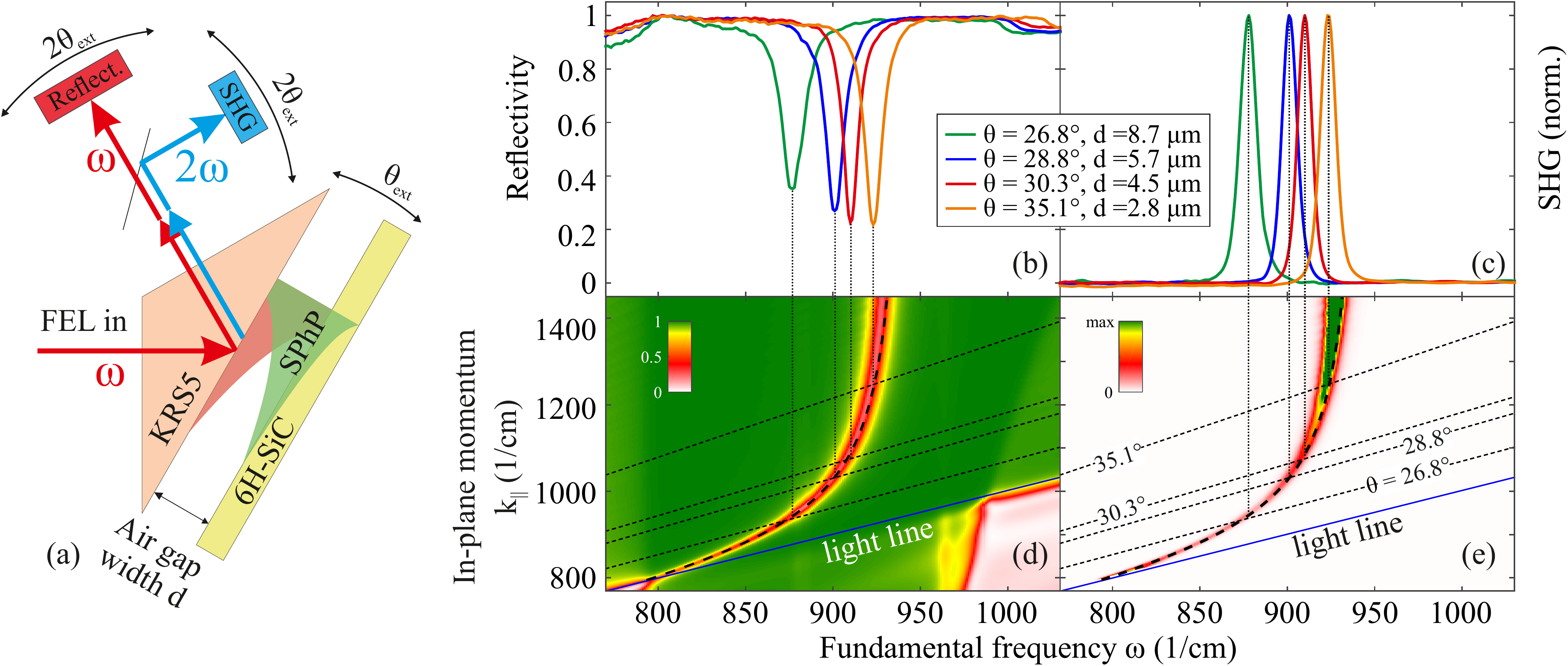}
\caption{(a) Schematic of the experimental setup (not to scale) for Otto-type prism coupling to propagating SPhPs. Tuning the FEL frequency $\omega$, air
 gap width $d$, and excitation-detection angle $\theta_\mathrm{ext}-2\theta_\mathrm{ext}$ allows for full control over the SPhP excitation conditions. 
We measure both SHG intensity and reflectivity of the fundamental beam simultaneously using a dichroic beam splitter.
For illustration, we also schematically show the evanescent fields leaking into the air gap from the prism side (red shaded) and SPhP (green shaded) at critical coupling conditions. Experimental reflectivity (b) and SHG (c) spectra taken for multiple incidence angles, each near the respective critical coupling gap width (see legend), leading to the most efficient SPhP excitation. Calculated reflectivity (d) and SHG (e) maps evaluated at critical coupling conditions show how the signals follow the SPhP dispersion (dot-dashed lines): for each given incidence angle $\theta$, the in-plane momentum of the fundamental light (sloped dashed lines) intersects with the SPhP dispersion at the respective spectral positions of the SPhP resonances in the reflectivity and SHG (vertical dotted lines).}
\label{fig:setup}
\end{figure*} 

Excitation of propagating SPhPs and detection of their reflectivity and SHG response is realized experimentally as schematically shown in Fig.~\ref{fig:setup} (a). We implement the Otto  arrangement\cite{Otto1968,Falge1973,Neuner2009} by placing a triangular prism (KRS5, $n_{\rm prism} \approx 2.4$, Korth) operated in the regime of total internal reflection onto a motorized mount in front of the sample, allowing for continuous tuning of the air gap width $d$. Rotation of the prism-sample assembly by angle $\Delta \theta_{ext}\approx n_{\rm prism} \Delta \theta$ changes the in-plane momentum $k_\parallel = \frac{\omega}{c} \, n_{\rm prism} \sin\theta$ of the incoming wave, allowing to excite SPhPs at different points along the dispersion.\cite{Falge1973} Here, $\theta$ is the incidence angle inside the prism of refractive index $n_{\rm prism}$, and $\omega$ the frequency of the incoming mid-IR beam. We use an infrared FEL\cite{Schollkopf2015} as tunable, narrow band p-polarized excitation source, and the reflected fundamental and SHG beams are detected after a dichroic beam splitter. As a sample, we use a semi-insulating 6H-SiC c-cut single crystal, see Suppl. Mat. for further details on the experiment.

Notably, the efficiency of coupling the incoming light to the SPhPs in this geometry sensitively depends on the air gap width $d$.\cite{Neuner2009} The decay length $L$ of the evanescent waves into the air gap for both, the totally reflected incoming light and the SPhP, strongly varies with the in-plane momentum $k_\parallel$:\cite{Raether1988}
\begin{equation}
 L = \frac{\lambda}{2\pi\sqrt{k_\parallel^2c^2/\omega^2 - 1}},
 \label{eq:eva}
 \end{equation}
where $\lambda$ is the wavelength and $c$ the speed of light in vacuum. For small gaps $d \ll L$ with large overlap of the two evanescent waves, the strong radiative coupling of the SPhP back into the prism is a significant loss channel and prevents efficient excitation, while for large gaps $d \gg L$ the small overlap between the two evanescent waves inhibits an efficient energy transfer. There is, however, a critical coupling gap width $d_{\rm crit}$ where the radiative and intrinsic losses of the SPhP exactly balance each other, and critical coupling to SPhPs is achieved.\cite{Neuner2009}  This is illustrated with the red and green shaded areas in Fig.~\ref{fig:setup} (a) for prism-side and SPhP waves, respectively. Since  $k_\parallel$ strongly varies along the SPhP dispersion, both $L$ and the critical gap width $d_{\rm crit}$ also vary from less than 1~$\mu$m for large momenta to tens of $\mu$m at small momenta when approaching the light line.

In consequence, the Otto arrangement allows for a high-precision, direct measurement of the full surface polariton dispersion if the critical coupling conditions are adapted appropriately. 
This mapping of the SPhP dispersion is demonstrated in Fig.~\ref{fig:setup} (b) and (c) where we show experimental reflectivity and SHG spectra, respectively, at four different incidence angles, each taken near the respective $d_{\rm crit}$. Under these conditions, a reflectivity dip of $\approx 80\%$ indicates efficient excitation of the SPhP. The spectral position of the dip follows the SPhP dispersion, see  Fig.~\ref{fig:setup} (d), as the incidence angle is changed. At the same time, the field enhancement associated with the efficiently excited SPhP results in significant increase of the SHG yield, as evidenced by the single, narrow peak in each of the SHG spectra (c).

For comparison, we show calculated reflectivity and SHG maps at critical coupling conditions in Fig.~\ref{fig:setup} (d) and (e), respectively, plotted as a function of fundamental frequency $\omega$ and in-plane momentum $k_\parallel$. The reflectivity is calculated using the transfer matrix approach,\cite{Berreman1972,Yeh1979,Xu2000} see Suppl. Mat. for details, while the SHG intensity is computed using: 
\begin{eqnarray}
\label{eq:SHG}
I(2\omega) & \propto & | ( T_b \vec{E}_{SiC}(2\omega)) \\ \nonumber
 & & \cdot \left ( \chi^{(2)}(-2\omega;\omega,\omega) \vec{E}^2_{SiC}(\omega)/\Delta k\right )  |^2.
\end{eqnarray}
Here, $\vec{E}_{\rm SiC}(\omega)$ is the local optical field on the SiC side of the SiC-air interface which we obtain from the transfer matrix approach, $\chi^{(2)}$ the nonlinear susceptibility of 6H-SiC,\cite{Paarmann2016} while $\Delta k$ accounts for the wave vector mismatch for SHG in reflection.\cite{Paarmann2015,Paarmann2016} Additionally, we explicitly account for the field coupling of the nonlinear polarization by projecting it onto $ T_b \vec E_{\rm SiC}(2\omega)$, with $\vec E_{\rm SiC}(2\omega)$ the local field of the respective mode at $2\omega$ and $2k_\parallel$ propagating from SiC back into the prism, and $T_b$ the transmission coefficient back to the prism, see Suppl. Mat. for details on the SHG calculations.

\begin{figure}[bth]
\includegraphics[width=1\columnwidth]{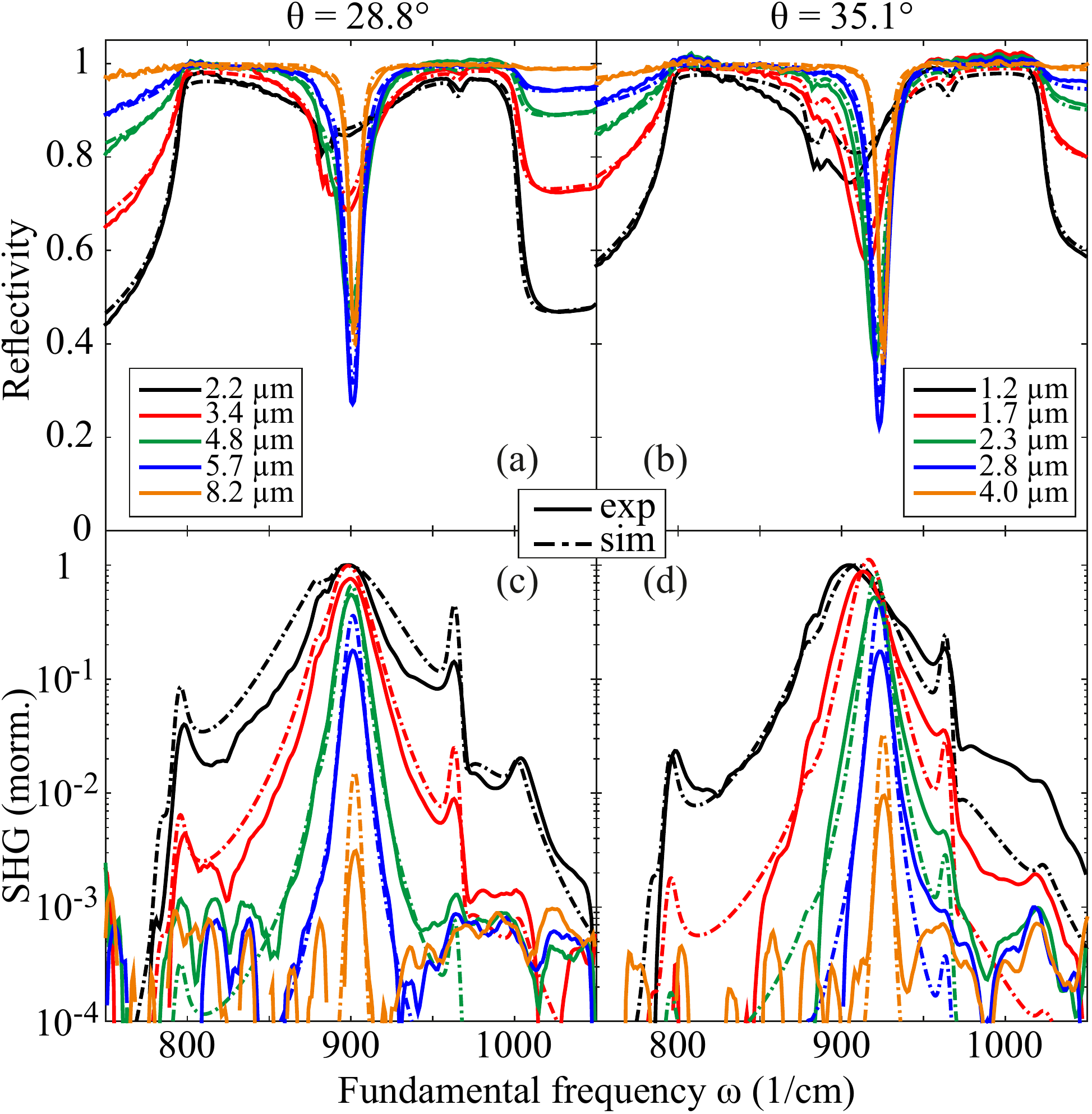}
\caption{Experimental (solid lines) and calculated (dot-dashed lines) reflectivity (a,b) and SHG (c,d) spectra of the 6H-SiC-air interface in the Otto geometry, shown for selected air gap widths $d$ (see legends) and incidence angles $\theta=28.8^\circ$ (a,c)  and $\theta=35.1^\circ$ (b,d). The SHG spectra were normalized to the maximum at the smallest gap width. Please note the logarithmic scale in (c,d).}
\label{fig:spectra}
\end{figure}

To demonstrate the critical behavior of the response, we acquired reflectivity spectra for multiple values of the air gap width $d$ for selected incidence angles, exemplified for $\theta = 28.8^\circ$ and $\theta = 35.1^\circ$ in Fig.~\ref{fig:spectra}~(a) and (b), respectively. At the smallest gap (black lines), a shallow broad dip in the reflectivity at $\omega \approx 900$~cm$^{-1}$ reports on excitation of highly lossy SPhPs with pronounced radiative coupling. As $d$ is increased, the reflectivity dip increases in amplitude and narrows down while simultaneously blue-shifting. A maximum dip depth of $\approx 80\%$ (blue lines) indicates that the critical coupling conditions are reached.
For even larger gaps (orange lines), the amplitude of the dip drops again, while the narrowing and blue-shifting of the resonance converge, approaching the intrinsic line width and frequency of the uncoupled surface polariton, cf. Fig.~\ref{fig:setup}~(d). The qualitative behavior is identical between the two incidence angles shown here. However, the respective air gaps are clearly different, see legends in (a,b), where $d$ essentially scales with the evanescent length $L$ in Eq.~\ref{eq:eva}. The calculated reflectivity perfectly reproduces the experimental data, including subtle features due to the 6H-SiC crystal anisotropy, such as reflectivity dips at the zone-folded weak modes at $\approx 885$~cm$^{-1}$ and the axial LO phonon at $964$~cm$^{-1}$.\cite{Engelbrecht1993,Bluet1999,Paarmann2016} 

The simultaneously acquired SHG spectra are shown in Fig.~\ref{fig:spectra} (c,d). At small gaps, the data exhibit some additional spectral features,\cite{Paarmann2015,Paarmann2016} however, a pronounced SPhP resonance can be clearly distinguished (please note the logarithmic scale), with SPhP peak positions and line widths qualitatively following those observed in the reflectivity as $d$ is increased. However, the amplitude behavior is clearly different for the SHG; we observe a steady decrease of the SHG amplitude at the SPhP resonance with increasing $d$.  The calculated SHG spectra also shown in Fig.~\ref{fig:spectra} (c,d, dash-dotted lines) not only reproduce all the features in the spectra with high accuracy. They also predict the steady decrease of SPhP resonance amplitude of the SHG with increasing width $d$ of the air gap. This is very surprising, since the maximum field enhancement and, in consequence, most efficient nonlinear signal generation is expected at critical coupling.\cite{Raether1988,Neuner2009}

\begin{figure}[bth]
\includegraphics[width=\columnwidth]{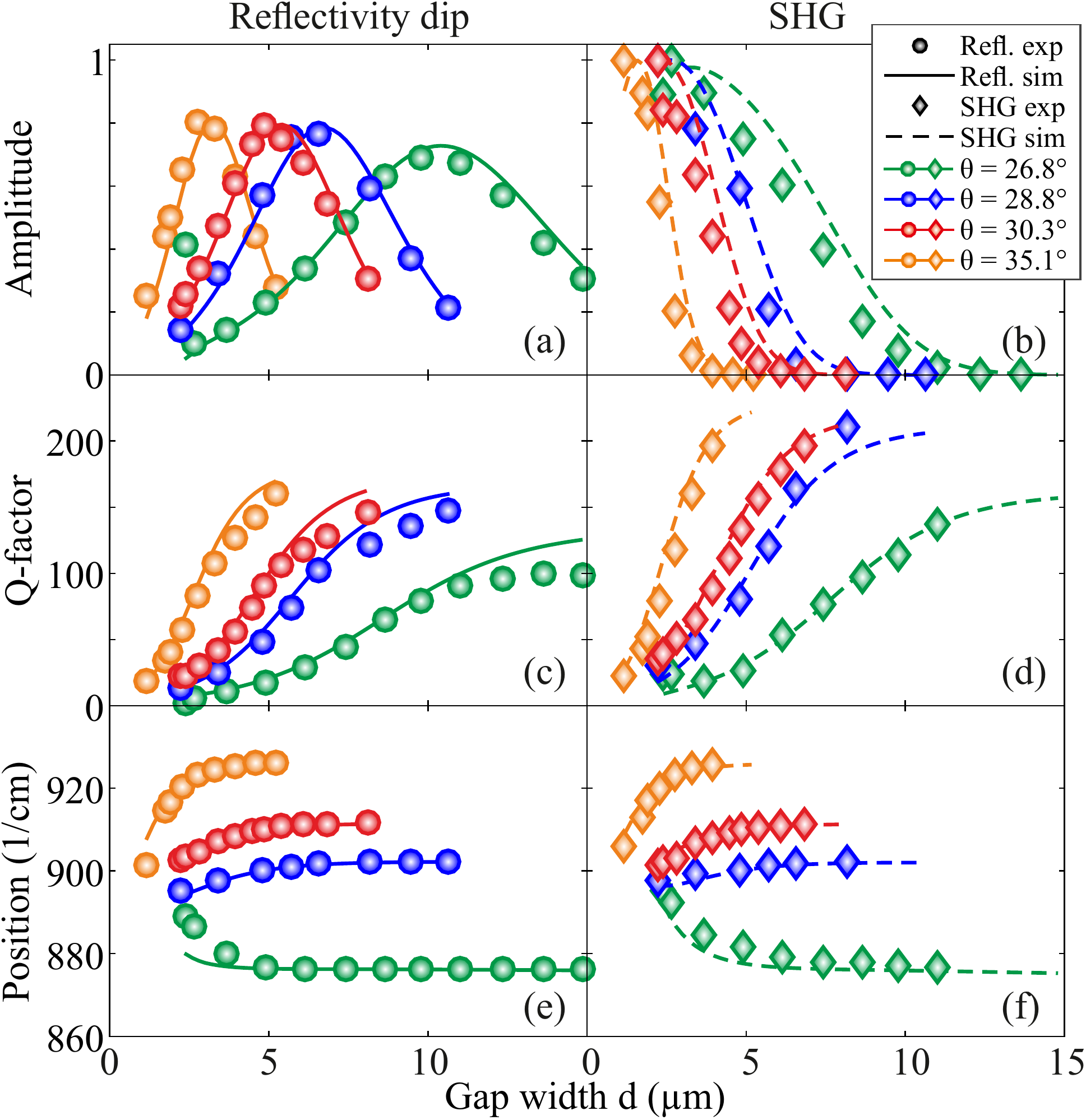}
\caption{Air gap width $d$-dependence of the SPhP resonance amplitude (a,b), Q-factor (c,d), and spectral position (e,f) at four different incidence angles, as extracted from the experimental (symbols) and calculated (lines) reflectivity (a,c,e) and SHG (b,d,f) spectra. Critical coupling is achieved at the respective maximum of the reflectivity dip amplitude in (a). The Q-factors increase and the spectral positions shift as $d$ is increased, converging towards the values of the weakly coupled, intrinsic surface polariton resonance, more rapidly with increasing $\theta$ and $k_\parallel$, i.e., further away from the light line.} 
\label{fig:critical}
\end{figure}

We summarize the SPhP resonance behavior for the full experimental data set in Fig.~\ref{fig:critical}. Lorentzian fits of the SPhP resonance in the experimental (symbols) and theoretical (lines) reflectivity (a,c,e) and SHG  (b,d,f) spectra yield the amplitude (a,b), Q-factor (c,d), and position (e,f) of the reflectivity dip and SHG peak, respectively. The high Q-factors, i.e., the ratios between center frequency and line width, mark the high quality of the SPhP resonance. Notably, the SHG resonances are intrinsically narrower as compared to the reflectivity dips due to the second-order nature of the interaction, leading to the consistently higher Q-factors. Remarkable agreement between experiment and theory is observed throughout the full data set. We note here that these calculations use a single set of parameters globally extracted from the full reflectivity data set, see Supp. Mat. for details.

The direct comparison of the SPhP resonance amplitudes of the reflectivity dip and the SHG peak, see Fig.~\ref{fig:critical} (a) and (b), respectively, for the different incident angles suggest a correlation between the critical coupling gap marking the strongest modulation of the reflectivity and the apparent decay length of the SHG amplitude. To understand these observations, we need to consider two mechanisms determining the detectable SHG signal in the far field: (i) efficiency of SHG due to the local field enhancement provided by the SPhPs, and (ii) the out-coupling of that nonlinear signal across the air gap into the prism, and into the far field. In the case of low losses, the local field enhancement associated with surface polariton excitations is typically expected to follow the magnitude of the reflectivity dip,\cite{Raether1988,Neuner2009}  i.e., the maximum enhancement should be achieved at the critical coupling condition. 

It is, however, important to realize that the SHG is generated with large in-plane momentum $k_{\rm \parallel, SHG} = 2 k_{\rm SPhP} > k_{\rm 0,SHG}$, where $k_{\rm 0,SHG}$ is the wavenumber of the SHG light propagating in air. This corresponds to the condition of total internal reflection for the SHG at the SiC-air interface, and only an evanescent wave is expected to leak into the air gap. Evaluating Eq.~\ref{eq:eva} for this wave reveals exactly half the evanescent length for the SHG as compared to the fundamental radiation on both sides of the air gap, resulting in an extremely poor overlap and largely suppressed far field coupling of the SHG at critical coupling conditions. In our SHG calculations, this effect is explicitly accounted for by the gap width $d$-dependence of $ T_b \vec{E}_{\rm SiC}(2\omega)$ in Eq.~\ref{eq:SHG}, whose amplitude rapidly decays with $d$. In fact, we find that at critical coupling only $\sim 0.1\%$ of the generated nonlinear signal is harvested into the far-field intensity. 

\begin{figure*}[bth]
\includegraphics[width=1\textwidth]{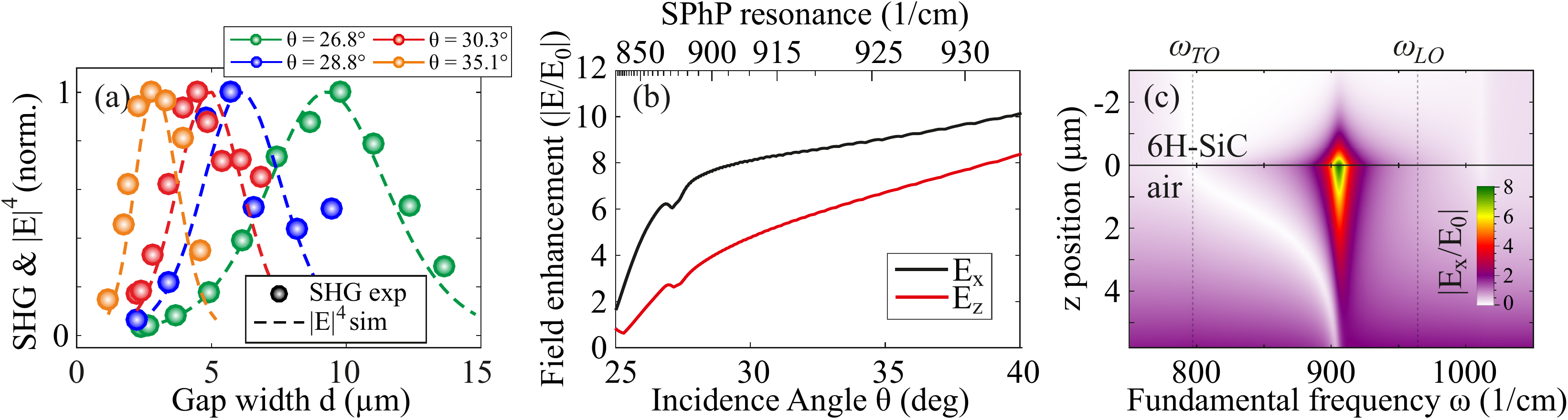}
\caption{(a) The SHG signals corrected for out-coupling losses (symbols) perfectly follow the fourth power of the calculated local optical fields (lines). (b) SPhP optical field enhancement $|E/E_0|$ with $E_0$ the incoming field magnitude, evaluated at critical coupling gap for in-plane $E_x$ (black) and out-of-plane $E_z$ (red) field components as a function of internal incidence angle (bottom) and corresponding SPhP resonance position (top). (c) Example of the spatio-spectral distribution of in-plane field enhancement, shown at critical coupling for $\theta = 28.8^\circ$ with $d= 5.8~\mu m$. All electric fields are evaluated on the SiC side of the SiC-air interface.}
\label{fig:fieldenhancement}
\end{figure*}

Therefore, we can recover the second harmonic intensity generated in SiC by correcting the experimental signals for the out-coupling efficiency $|T_b \vec E_{\rm SiC}(2\omega)|^2$, as shown in Fig.~\ref{fig:fieldenhancement}~(a). As described by the second factor in Eq.~\ref{eq:SHG}, these signals now follow fourth power of the resonantly enhanced fundamental optical fields $|E_{\rm SiC}(\omega)|$, which are also plotted there. With this information, we can confidently extract the magnitude of the field enhancement in SiC at critical coupling, which is plotted in Fig.~\ref{fig:fieldenhancement}~(b) along the SPhP dispersion. We plot both in-plane and out-of-plane fields, $E_x$ and $E_z$, respectively, since due to the symmetry of the $\chi^{(2)}$ tensor, both components contribute significantly to the SHG output.\cite{Paarmann2016}  Notably, the small dip in these curves at $\theta \approx 27.5^\circ$ marks the resonant interaction of the SPhP with the zone-folded phonon modes in 6H-SiC.\cite{Razdolski2016} For illustration, we also show an example of the spatio-spectral distribution of field enhancement along the normal-to-surface direction $z$ in Fig.~\ref{fig:fieldenhancement} (c). 

The excellent agreement between the experimental and calculated SHG response corroborates our observation of the extremely efficient nonlinear-optical conversion inside the SiC crystal, in particular in comparison to SPPs.  From our data, we estimate an exceptional nonlinear conversion efficiency exceeding $\sim 10^{-6}$, which is a result of (i) the large field enhancement due to the high Q-factor of the SPhP resonance, and (ii) the broken inversion symmetry as well as proximity to an ionic resonance of the nonlinear susceptibility\cite{Mayer1986,Paarmann2016} in the SPhP host. 

The latter two features, broken inversion and ionic resonance, are unique fingerprints of SPhP materials as opposed to noble metals used in plasmonics.\cite{Maier2007} Our approach is applicable to all non-centrosymmetric, polar dielectrics, such as $\alpha$-quartz, ZnO, AlN, InP, GaAs, ZnTe, to name a few, known to exhibit large nonlinear coefficients, as well as artificially designed hybrid materials\cite{Caldwell2016} with yet unknown SPhP dispersion relation. Additionally, the strong dispersion of SPhPs provides spectral tunability of the field enhancement resonance with extraordinarily high Q-factors. 

The combination of these features suggests several appealing scenarios for future applications of nonlinear nanophononics.\cite{Razdolski2016} Previous linear sensing schemes\cite{Neuner2010,Zheng2017} could be extended into the nonlinear domain with larger contrasts and higher Q-factors. Taking advantage of the large resonant, high-Q nonlinear signals from the broken-inversion SPhP host, these approaches could straight-forwardly be implemented using narrow-bandwidth quantum-cascade lasers. Similarly, our approach could be extended to four-wave mixing experiments,\cite{Palomba2008,Leon2014} sensitive to either the SPhP host or a liquid phase analyte in the gap, cf. Fig.~\ref{fig:fieldenhancement}~(c) for the relevant length scales of field enhancement inside the gap. This could, for instance, lead to a drastic resonant enhancement of time-resolved vibrational spectroscopy signals from molecules in solution.\cite{Hamm1995}  Additionally, we also note that SPhP materials are typically largely transparent above the reststrahl spectral range. Together with the generation of nonlinear signal under conditions of total internal reflection, this suggests that the Otto geometry could be used for high-contrast nonlinear waveguide coupling.\cite{Zheng2016} 

\section{Conclusion}

In conclusion, we demonstrated the first SHG from critically coupled SPhPs, enabling high-precision  measurement of the associated optical field enhancement, here shown for 6H-SiC as a model system. We use the critical coupling behavior of the Otto geometry to maximize the SPhP excitation efficiency. Despite poor far-field coupling of the SHG signals, which is shown to be a generic feature of the Otto arrangement, the large bulk nonlinearity of the crystal facilitates exceptional nonlinear-optical conversion from SPhPs, owing to the broken inversion symmetry of the  host material and the high quality of the SPhP resonance. Being applicable to a wide range of polar dielectrics supporting SPhPs, our approach could be employed to extract the unknown polariton dispersion and field enhancement in artificially designed SPhP hybrids, and opens the path to several new and appealing applications of nonlinear nanophononics. 
 
\section{Acknowledgments}
The authors thank  K. Horn (FHI Berlin) for providing the SiC sample.


\end{document}


\maketitle

\beginsupplement


\section{Experimental}

\begin{figure}[bht]
\includegraphics[width = \columnwidth]{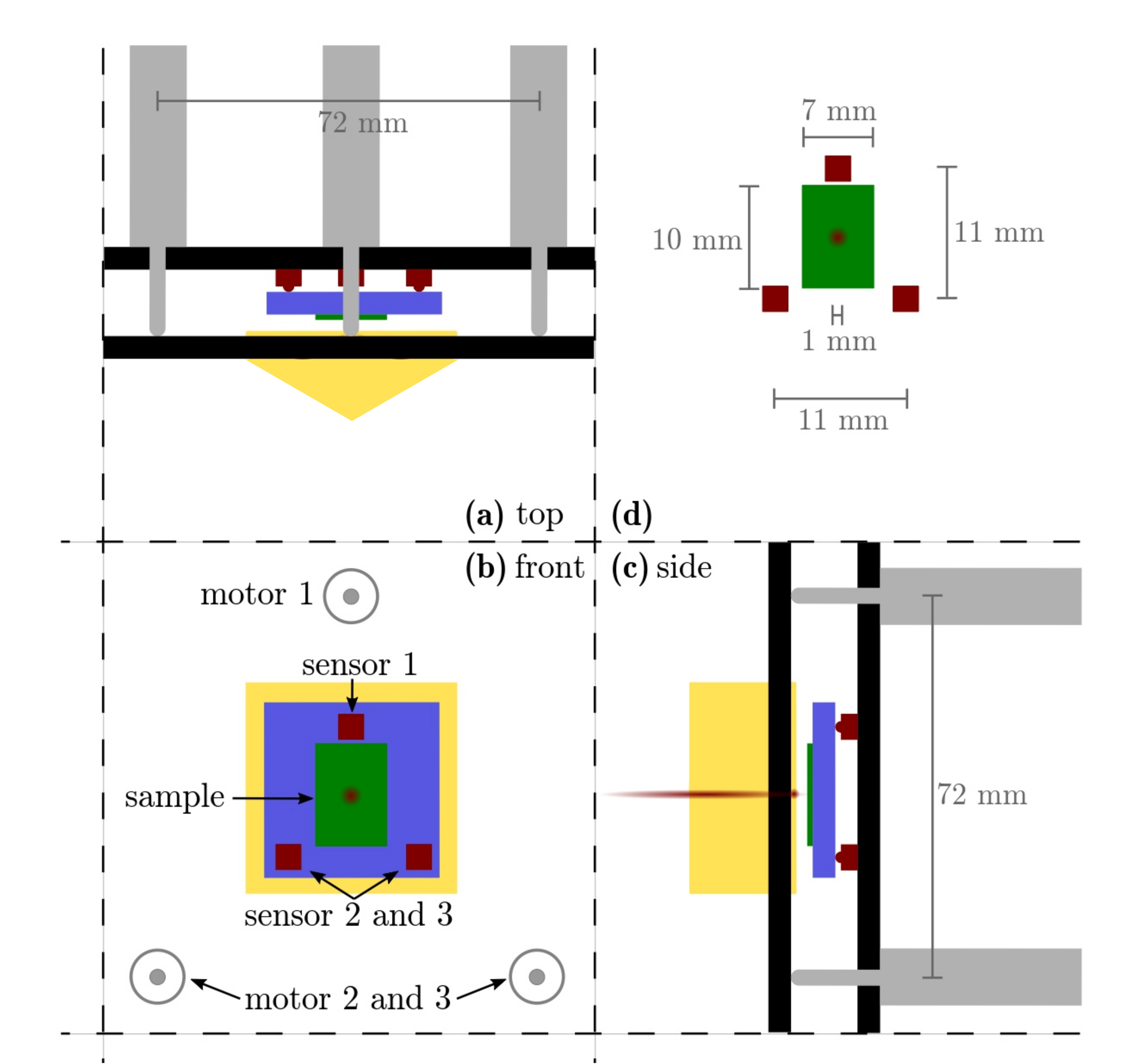}
\caption{Practical implementation of the Otto-type prism coupling to surface polaritons. Three motorized actuators push (against springs) the prism mount away from the sample. When pulled back, the point of contact is detected using three mN force sensors, allowing for parallel alignment of the prism and sample surfaces and continuous scanning of the gap width $d$.}
\label{sfig:setup}
\end{figure}

The experimental setup was schematically shown in Fig.~1 (a) of the main text. The Otto-type prism coupling is realized by placing a triangular prism (KRS5, 25mm high, 25mm wide, angles of 30$^\circ$, 30$^\circ$, and 120$^\circ$, Korth Kristalle GmbH) on a motorized mount in front of the sample. Details on the implemtation are shown in Fig.~\ref{sfig:setup}. Three motorized actuators (Newport TRA12PPD) are used to push (against springs) the prism away from the sample, which is mounted on a plate (25mm diameter) equipped with three mN force sensors (Honeywell FSS1500SSB). Each force sensor indicates the point of contact, allowing for parallel alignment of the sample and the prism back surface. The motorized actuators allow for continuous tuning of the air gap width $d$. 

We note that macroscopic protrusions and non-flatness of the sample can lead to strain between sample and prism, resulting in a lower limit of air gap width (typically $\approx 1.5~\mu$m) and slight nonlinearities of the actual air gap width and the actuator readback values, specifically  at close approach. Therefore, we fitted the values of $d$ from the reflectivity data, using the procedure described in Sec.~\ref{sec:fit}, the results of which are in qualitative (but not quantitative) agreement with the actuator readback values. Gap width scans of the signals are typically performed by starting at closest achievable approach, followed by defined steps (identical for all three actuators) away from the sample.

Given the large refractive index of KRS5 of $n_{KRS5} \approx 2.4$, external angles differ respectively when going away from $\theta_\mathrm{ext} = 30^\circ$, i.e., the internal angle $\theta$ inside the prism is $\theta = 30^\circ + \arcsin(\sin(\theta_\mathrm{ext}-30^\circ)/n_{KRS5})$. This arrangement allows for internal angles from below the total internal reflection cutoff at $\theta_\mathrm{cutoff} \approx 25^\circ$ to about $\theta = 52^\circ$, covering the majority of the SPhP dispersion. We note that using more convenient semi-cylindrical prisms results in strong horizontal focusing, increasing angular spread of the beam and thereby much degraded momentum resolution. Additionally, this horizontal focusing distorts the beam to an elliptical shape making it harder to focus onto small detectors.

We use an infrared FEL\cite{Schollkopf2015} as tunable, narrowband, and p-polarized excitation source. We focus the beam mildly ($f\approx60$~cm, Edmund Optics) onto the sample surface to minimize angular spread of the beam. Details on the IR-FEL are given elsewhere.\cite{Schollkopf2015} In short, the electron gun is operated at a micropulse repetition rate of 1~GHz with an electron macropulse duration of 10~$\mu$s and a  macropulse repetition rate of 10~Hz. The electron energy is set to 31~MeV, allowing to tune the FEL output wavelength between $\sim7-18$~$\mu$m ($\sim1400-550$~cm$^{-1}$) using the motorized undulator gap, providing ps-pulses with typical full-width-at-half-maximum (FWHM) of $<5$~cm$^{-1}$. 

Reflected fundamental and SHG beams are split using a $~7~\mu$m long-pass filter (LOT). The reflectivity is detected with a home-built pyro-electric sensor (Eltec 420M7). The SHG signal is detected with a liquid nitrogen cooled mercury cadmium telluride/indium antimony sandwich detector (Infrared Associates), after blocking the fundamental light using 5~mm of MgF$_2$ and a  $~7~\mu$m short-pass filter (LOT). Additionally, intrinsic harmonics in the FEL beam are blocked prior to the setup using a $~7~\mu$m and a $~9~\mu$m long-pass filter (both LOT). We also measure reference power on a single shot basis to minimize the impact of shot-to-shot fluctuations of the FEL using the reflection off a thin KBr plate placed in the beam prior to the setup. This reference signal is also detected by a pyro-electric detector, and we normalize all signals to this reference signal on a single-shot basis. Finally, we determine the spectral response function of the setup by measuring the reflectivity for very large gap width $d \approx 40~\mu$m, i.e. in total reflection across the full spectral range, for each incidence angle. All spectra are normalized for these reference data.

\begin{figure*}[th!]
\includegraphics[width=0.6\textwidth]{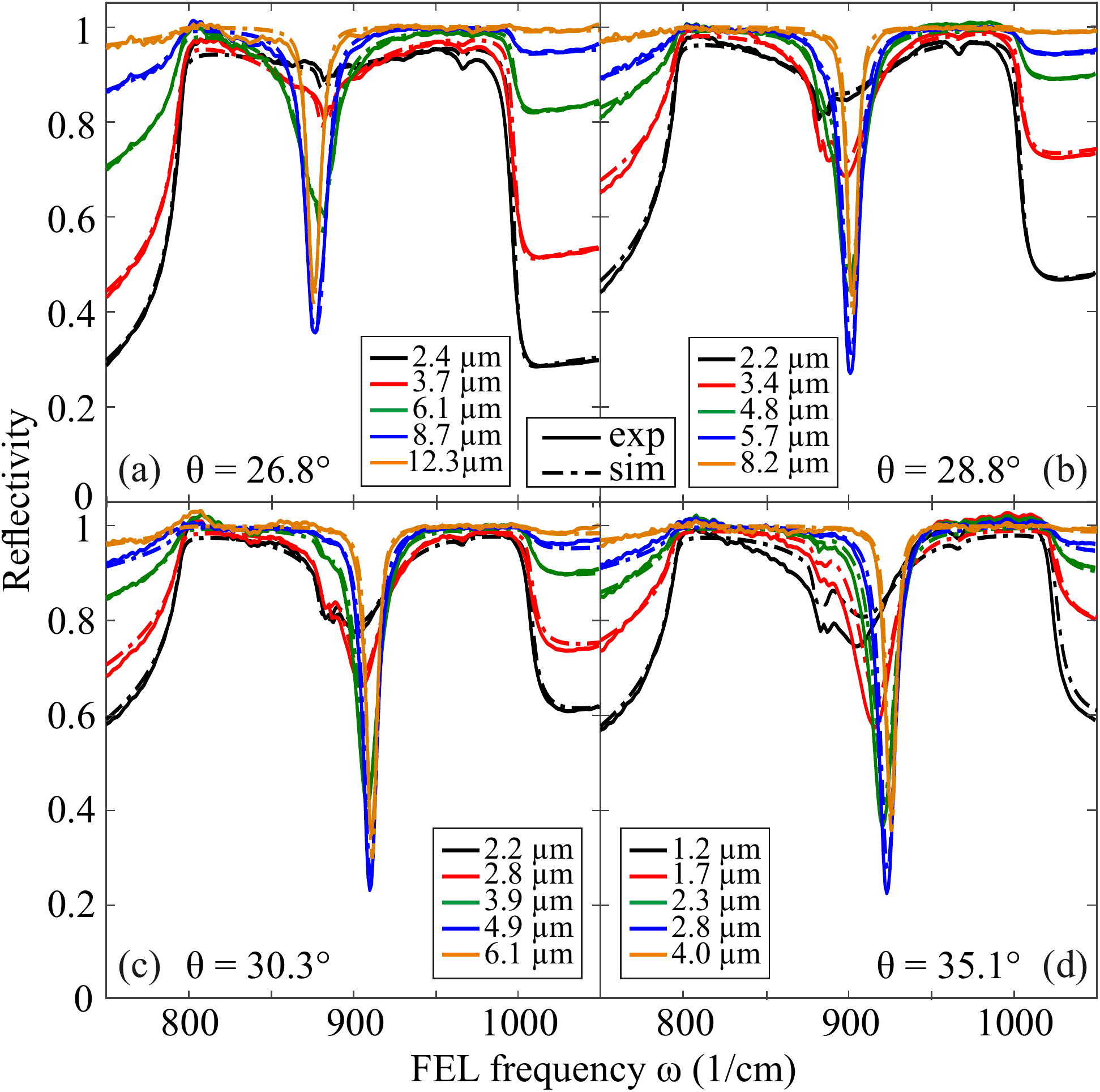}
\caption{Experimental (solid lines) and calculated (dot-dashed lines) reflectivity spectra of 6H-SiC in the Otto configuration, shown for four different incidence angles $\theta$ (a,b,c,d), each with a set of five selected air gap widths $d$ (see legends).} 
\label{fig:reflData}
\end{figure*}

\section{Transfer Matrix Approach}
\label{transfermatrix}

For all calculations of reflectivity, local fields, and transmittance, a $4\times 4$ - matrix formalism was developed capable of treating any number of layers, each described by a dielectric tensor $\hat{\epsilon}$. In contrast to other approaches,\cite{Berreman1972,Yeh1979,Lin-Chung1984} our formalism is numerically stable and handles isotropic and anisotropic materials simultaneously, including the substrate. 

The multilayer surfaces are chosen to extend perpendicular to $z$ and the incident wave impinges in the $xz$-plane with wavevector $\kvec_i=\frac{\omega}{c}(\xi,0,q_i)$ in layer $i$. The $x$-component $\xi=\sqrt{\epsilon_{inc}} \sin(\theta)$ is conserved throughout the complete multilayer system, and $q_i$ is the unit-free, layer-dependent $z$-component.  In any medium, Maxwell's equations yield four independent plane-wave solutions characterized by $q_{ij}$ ($j=1..4$), which are determined as shown by Berreman.\cite{Berreman1972} In order to avoid instability and discontinuities, the four solutions have to be ordered unambiguously, which is done based on the work of \etal{Li}.\cite{Li1988}

In continuation, the treatment of singularities of \etal{Xu} \cite{Xu2000} is employed in order to generalize the formalism to any isotropic or anisotropic medium, which yields the electric field components $\Evec_{ij}=(\gamma_{ij1}, \gamma_{ij2}, \gamma_{ij3})$ of the four eigenmodes $j$ in layer $i$. With these components, a matrix \AMa{i}~is formulated (see equation 24 in ref. \cite{Xu2000}), representing Maxwell's boundary conditions at the interface $i-1$, $i$. The \Evec-fields are then connected by the interface matrix \LMa{i}~as follows: 
\begin{equation}
\Evec_{i-1}=\AMa{i-1}^{-1}\AMa{i}\Evec_{i} =: \LMa{i}\Evec_{i}.
\end{equation}
The propagation through layer $i$ is described by a diagonal matrix \PMa{i}~with elements $e^{-i \frac{\omega}{c} q_{ij}d_i}$, where $j=1..4$ is the nth diagonal element. The transfer matrix \FullTMa{N}~of the complete multilayer system reads
\begin{equation}
\FullTMa{N}=\LMa{1} \PMa{1} \LMa{2} \PMa{2} ... \LMa{N} \PMa{N} \LMa{N+1}.
\label{FullTMa}
\end{equation}
Finally, reflection and transmission coefficients for $p$-polarization are calculated in terms of \FullTMa{N}~components as shown by Yeh \cite{Yeh1979}.


\section{Fitting Procedure}
\label{sec:fit}
The reflectivity data were fitted employing the matrix formalism, and convolving the results to account for finite broadening of the incidence angle and FEL frequency, $\Delta\theta=0.2^{\circ}$ and $\Delta\omega=4.4 cm^{-1}$, respectively. For each incidence angle, the data were normalized by a reference spectrum taken at large gap width $d$ ($\approx 40 \mu m$). This accounts for the spectral response function of the setup, while additionally slow FEL power drifts are accounted for in the fitting procedure by a multiplication factor $a$ for every individual spectrum ($0.98<a<1.04$). 

The data sets of all four incidence angles $\theta$ were fitted globally, where $\theta$, $\Delta\theta$, $\Delta\omega$, and the 6H-SiC parameters such as TO and LO frequencies and damping were global fitting parameters, while gap widths $d$ and factors $a$ were adjusted individually. Thereby, $\Delta\theta$ was fitted only for the two smaller angles, and $\Delta\omega$ for the other two. This is justified due to the shape of the SPhP dispersion, being more sensitive to the incidence angle but less to the frequency at smaller $\theta$, and vice versa for larger angles. 

By fitting all four data sets in several iterations, the global parameters converged common values. In a final run, all data sets were fitted again using averages of the global parameters, yielding the curves and parameters shown in fig. \ref{fig:reflData}.


\section{Calculation of the Second Harmonic Intensity}

The calculations of the second harmonic intensity are largely based on previous work of SHG spectroscopy using free space excitation.\cite{Paarmann2016} In order to correctly describe the multilayer (KRS5/air/SiC) system, we needed to replace the Fresnel transmission coefficients with  appropriate expressions derived from the transfer matrix approach. Particular care is needed for the transmission of the outgoing SHG, as will be discussed below.

We first calculate the nonlinear polarization $\vec P^{NL}$ at $2\omega$ generated in SiC, before projecting it onto the appropriate mode at $2\omega$ responsible for the out-coupling of the SHG into the far field. Therefore, we rewrite Eq.~2 of the main text as:
\begin{equation}
\label{eq:SHG}
I(2\omega)  \propto  | ( T_b \vec{E}_{SiC}(2\omega)) \cdot \vec{P}^{NL} (2\omega) /\Delta k |^2
\end{equation}
with 
\begin{equation}
\label{P_NL}
\vec{P}^{NL}(2\omega) \propto ( \chi^{(2)}(-2\omega;\omega,\omega) \vec{E}^2_{SiC}(\omega) )  |^2.
 \end{equation}
We use the dispersion of the nonlinear susceptibility tensor $\chi^{(2)}(-2\omega;\omega,\omega)$ of 6H-SiC as extracted in our previous work.\cite{Paarmann2015,Paarmann2016} The local fields $\vec{E}^2_{SiC}(\omega)$ on the SiC side of the air-SiC interface are calculated as follows. 

First, we evaluate the transfer matrix $\FullTMa{}$ for a specific configuration of incidence angle, frequency and air gap width, employing the algorithm presented in Sec.~\ref{transfermatrix}. Then, the in-plane field component $E_{x,SiC}$ can be directly calculated from the transfer matrix $\FullTMa{}$ as the transmission coefficient for p-polarized light:\cite{Yeh1979}
\begin{equation}
\label{ExSiC}
E_{x,SiC} = \frac{\FullTMa{} [3,3]}{\FullTMa{} [1,1]\FullTMa{} [3,3]- \FullTMa{} [1,3]\FullTMa{} [3,1]}
\end{equation}
where $\FullTMa{} [i,j]$ denote the $(i,j)$ components of the $4\times4$ matrix (Eq.~\ref{FullTMa}). The out-of-plane component is determined using Maxwell's equation $ \bigtriangledown \times H = \epsilon \frac{dE}{dt}$:
\begin{equation}
\label{EzSiC}
E_{z,SiC} = -\frac{q_x \epsilon_x}{q_z \epsilon_z} E_{x,SiC}.
\end{equation}
with $q_{x(z)}$ and $\epsilon_{x(z)}$ the in-plane (out-of-plane) components of the wave vector and diagonal dielectric tensor of SiC, respectively. With this information, Eq.~\ref{P_NL} can now be evaluated. 

The out-coupling of the SHG is accounted for as follows. First, we calculate the transfer matrix $\FullTMa{}$ at $2\omega$ and, importantly, also $2k_\parallel$. Then, we invert $\FullTMa{} (2\omega)$ to describe the backward propagation. Finally we project $\vec P^{NL}$ onto the respective mode propagating from SiC back into the prism, and calculate the SHG intensity:
\begin{eqnarray} \nonumber
I(2\omega) &=& \frac{1}{|\Delta k|^2} \big|\left(\vec E_{SiC}(2\omega)	 \cdot \vec P^{NL}(2\omega)\right) \times \\ 
&& \times \frac{|E_{prism}(2\omega)|}{|E_{SiC}(2\omega)|}|^2
\end{eqnarray}
where $E_{x,SiC}(2\omega)\equiv 1$ and $E_{z,SiC}(2\omega)$ is calculated using the expression analogous to Eq.~\ref{EzSiC}. The second factor represents the transmission coefficient $T_b$ for the second harmonic field across the air gap. $E_{x,prism}(2\omega)$ is evaluated analogous to Eq.~\ref{ExSiC}, however, employing $\left [\FullTMa{} (2\omega)\right ]^{-1}$ as the transfer matrix, and $E_{z,prism}(2\omega)$ is again calculated using Eq.~\ref{EzSiC}.  
